\documentclass[sort&compress]{appolb}
\usepackage{graphicx}

\usepackage{amssymb}
\usepackage{amsmath}
\usepackage[usenames,dvipsnames]{color}
\definecolor{darkblue}{RGB}{0,0,196}
\definecolor{darkgreen}{RGB}{0,120,0}
\usepackage[colorlinks=true,linkcolor=darkblue,citecolor=darkblue,urlcolor=darkblue]{hyperref}

\def\ba{\begin{eqnarray}}
\def\ea{\end{eqnarray}}

\def\be{\begin{equation}}
\def\ee{\end{equation}}

\newcommand*\barM{{\overline{\cal M}\hspace{0.5mm}}} 

\begin{document}

\eqsec  

\title{The non-equilibrium attractor: \\Beyond hydrodynamics%
\thanks{Presented at the XXV Cracow EPIPHANY Conference on Advances in Heavy Ion Physics}%
}
\author{Michael Strickland
\address{Department of Physics, Kent State University, Kent, OH 44242 USA}
}
\maketitle

\begin{abstract}
The quark-gluon plasma created in heavy-ion collisions is not in local thermal equilibrium at early times.  Despite this, dissipative hydrodynamics describes the evolution of the energy-momentum tensor quite well after only roughly \mbox{0.5 - 1 fm/c}.  This can be understood using the concept of a non-equilibrium dynamical attractor.  The attractor is a uniquely identifiable solution to the dynamical equations to which all solutions are drawn as the system evolves.  Once solutions collapse onto the non-equilibrium attractor they are ``pseudo-thermalized'' in the sense that they have lost information about the precise initial conditions used, but are not yet in exact local thermal equilibrium.  Here I review recent work which demonstrates that there exists a non-equilibrium attractor in full kinetic theory models which goes beyond the usual low-order momentum moments considered in hydrodynamical treatments.
\end{abstract}

  
\section{Introduction}

At the highest collision energies the Large Hadron Collider (LHC) collides lead (Pb) nuclei with nucleon-nucleon center-of-mass energy $\sqrt{s_{NN}} = 5.5$ TeV.  At these energies, central Pb-Pb collisions create a quark-gluon plasma (QGP) that has an initial central temperature of $T_0 \simeq $ 600 MeV at $\tau_0 = 0.25$ fm/c after the nuclear pass through \cite{Alqahtani:2017jwl,Alqahtani:2017tnq}.  The QGP then expands and cools, eventually hadronizing into the particles which are detected experimentally.  The evolution and hadronization of the QGP created in such events has been modeled quite successfully using relativistic dissipative hydrodynamics \cite{Averbeck:2015jja,Jeon:2016uym,Romatschke:2017ejr,Florkowski:2017olj,Alqahtani:2017mhy}.  Observables such as the relative yields of different hadrons, identified hadronic spectra, and elliptic flow have been well reproduced by hydrodynamic models with a shear viscosity to entropy density ratio $\eta/s \simeq 0.2$.

Since traditional viscous hydrodynamics (vHydro) approaches rely on a linearization around isotropic thermal equilibrium, the conventional wisdom was that the phenomenological success of vHydro implied that the QGP created in heavy ion collisions was in or close to being in isotropic thermal equilibrium for its entire lifetime.  In a challenge to this conventional wisdom, calculations of thermalization and isotropization of the QGP using both weak and strong coupling methods found that there were sizable non-equilibrium corrections at early times and, in particular, a large early-time pressure anisotropy in the local rest frame, ${\cal P}_L \ll {\cal P}_T$ \cite{Strickland:2013uga}.  Despite these large pressure anisotropies, it was found that after a certain time, called the {\em hydrodynamization time} $\tau_{\rm hydro}$, the evolution obtained from both strongly and weakly coupled descriptions could be accurately modeled using dissipative relativistic hydrodynamics \cite{Chesler:2008hg,Heller:2013oxa,Keegan:2015avk,Romatschke:2017vte,Strickland:2017kux}.  These studies found $\tau_{\rm hydro} \sim 2/T$, which at the highest LHC energies translates into \mbox{$\tau_{\rm hydro} \sim 0.5$ fm/c} when considering the center of the fireball created in a zero impact parameter collision.

The success of dissipative hydrodynamical evolution in the presence of large non-equilibrium deviations can be understood in the context of the {\em hydrodynamical attractor} \cite{Heller:2015dha}.  In 0+1d conformal viscous hydrodynamics, for example, one can reduce the hydrodynamical equations of motion to a single ordinary differential equation which, subject to the correct boundary conditions, provides a universal ``attractor'' solution.  If one solves the hydrodynamic equations with different initial conditions and plots the results versus $\bar{w} = \tau/\tau_{\rm eq}$, one finds that the solutions with different initial conditions converge to the universal attractor solution on a very short time scale (in the sense of small $\bar{w}$).  Beyond second-order viscous hydrodynamics, the existence of a non-equilibrium attractor has been demonstrated using numerical solutions to Einstein's equations obtained in the strong coupling limit of ${\cal N}=4$ supersymmetric Yang-Mills \cite{Heller:2013oxa,Keegan:2015avk,Romatschke:2017vte}, QCD effective kinetic theory simulations \cite{Keegan:2015avk,Kurkela:2015qoa,Romatschke:2017vte}, third-order viscous hydrodynamics \cite{JaiswalForth}, anisotropic hydrodynamics \cite{Strickland:2017kux}, and exact solutions to the Boltzmann equation in relaxation time approximation (RTA) subject to both Bjorken and Gubser flows~\cite{Romatschke:2017vte,Strickland:2017kux,Behtash:2017wqg,Behtash:2018moe,Denicol:2018pak,Strickland:2018ayk,Behtash:2019txb}.   

In this proceedings, I report on work contained in Ref.~\cite{Strickland:2018ayk}.  Therein, I demonstrated that, using kinetic theory, the idea of the non-equilibrium attractor can be extended beyond the low-order moments of the one-particle distribution function typically considered in hydrodynamic approaches.  This was done by considering the evolution of general moments of the one-particle distribution function using an exact solution of the conformal 0+1d Boltzmann equation in RTA~\cite{Florkowski:2013lza,Florkowski:2013lya}.  This exact solution takes the form of a one-dimensional integral equation for the effective temperature which can be solved iteratively.  Once the effective temperature is known, one can solve for all moments of the one-particle distribution function and the distribution function itself.

\section{Exact solution for an arbitrary moment}

Herein I present results of exact solution to the RTA Boltzmann equation subject to boost-invariant and transversally homogenous Bjorken flow.  The underlying kinetic equation is simple
\be
p^\mu \partial_\mu  f(x,p) =  \frac{p \cdot u}{\tau_{\rm eq}(\tau)} \left( f_{\rm eq}-f \right) , 
\label{kineq}
\ee
where $\tau_{\rm eq}(\tau) = 5\bar\eta(\tau)/T(\tau)$ is the relaxation time with $\bar\eta(\tau)=\eta(\tau)/s(\tau)$ being the shear viscosity to entropy density ratio and $T(\tau)$ being the local effective temperature.  Eq.~\eqref{kineq} can be cast into simpler form by writing it in terms of manifestly boost-invariant variables~\cite{Bialas:1984wv,Bialas:1987en}.  The resulting simpler equation can easily be shown to have a general solution given by~\cite{Baym:1984np,Florkowski:2013lza,Florkowski:2013lya}
\be
f(\tau,w,p_T) = D(\tau,\tau_0) f_0(w,p_T) 
+  \int_{\tau_0}^\tau \frac{d\tau^\prime}{\tau_{\rm eq}(\tau^\prime)} \, D(\tau,\tau^\prime) \, 
f_{\rm eq}(\tau^\prime,w,p_T) \, ,  
\label{eq:exactsolf}
\ee
where $D$ is the damping function
\be
D(\tau_2,\tau_1) = \exp\left[-\int\limits_{\tau_1}^{\tau_2}
\frac{d\tau^{\prime\prime}}{\tau_{\rm eq}(\tau^{\prime\prime})} \right] .
\ee

Equation~\eqref{eq:exactsolf} can be turned into an infinite tower of equations for moments of the one-particle distribution function
\be
{\cal M}^{nm}[f] \equiv \int dP \,(p \cdot u)^n \, (p \cdot z)^{2m} \, f(\tau,w,p_T) \, ,
\label{eq:genmom1}
\ee
with the result being~\cite{Strickland:2017kux}
\ba
{\cal M}^{nm}(\tau) &=& \frac{\Gamma(n+2m+2)}{(2\pi)^2} \Bigg[ D(\tau,\tau_0) T_0^{n+2m+2} \frac{{\cal H}^{nm}\!\left( \tfrac{\alpha_0 \tau_0}{\tau} \right)}{[{\cal H}^{20}(\alpha_0)/2]^{(n+2m+2)/4}} \nonumber \\
&& \hspace{1cm} 
+ \int_{\tau_0}^\tau \frac{d\tau^\prime}{\tau_{\rm eq}(\tau^\prime)} \, D(\tau,\tau^\prime) \, 
T^{n+2m+2}(\tau') {\cal H}^{nm} \hspace{-1mm} \left( \tfrac{\tau'}{\tau} \right) \Bigg] ,
\label{eq:meqfinal}
\ea
with
\be
{\cal H}^{nm}(y) = \tfrac{2y^{2m+1}}{2m+1}  {}_2F_1(\tfrac{1}{2}+m,\tfrac{1-n}{2};\tfrac{3}{2}+m;1-y^2)  \, .
\ee
Note that certain moments map to familiar hydrodynamics variables, e.g. $n = {\cal M}^{10}$, $\varepsilon = {\cal M}^{20}$, and $P_L = {\cal M}^{01}$.

One can obtain a closed integral equation for $T(\tau)$ by consider the integral equation obeyed by ${\cal M}^{20} = \varepsilon = \varepsilon_{\rm eq}$ which simplifies to
\be
T^4(\tau) = D(\tau,\tau_0) T_0^4 \frac{{\cal H}\!\left( \frac{\alpha_0 \tau_0}{\tau} \right)}{{\cal H}(\alpha_0)} \nonumber \\
+ \int_{\tau_0}^\tau \frac{d\tau^\prime}{2 \tau_{\rm eq}(\tau^\prime)} \, D(\tau,\tau^\prime) \, 
T^4(\tau') {\cal H}\hspace{-1mm} \left( \frac{\tau'}{\tau} \right) . 
\label{t4eq}
\ee
This equation can be numerically solved iteratively~\cite{Florkowski:2013lza,Florkowski:2013lya}.  Once the solution for $T(\tau)$ is obtained, one can use this to solve for all other moments ${\cal M}^{nm}(\tau)$ using Eq.~\eqref{eq:meqfinal} and the full distribution function itself using Eq.~\eqref{eq:exactsolf}.

\begin{figure*}[t!]
\centerline{
\includegraphics[width=1\linewidth]{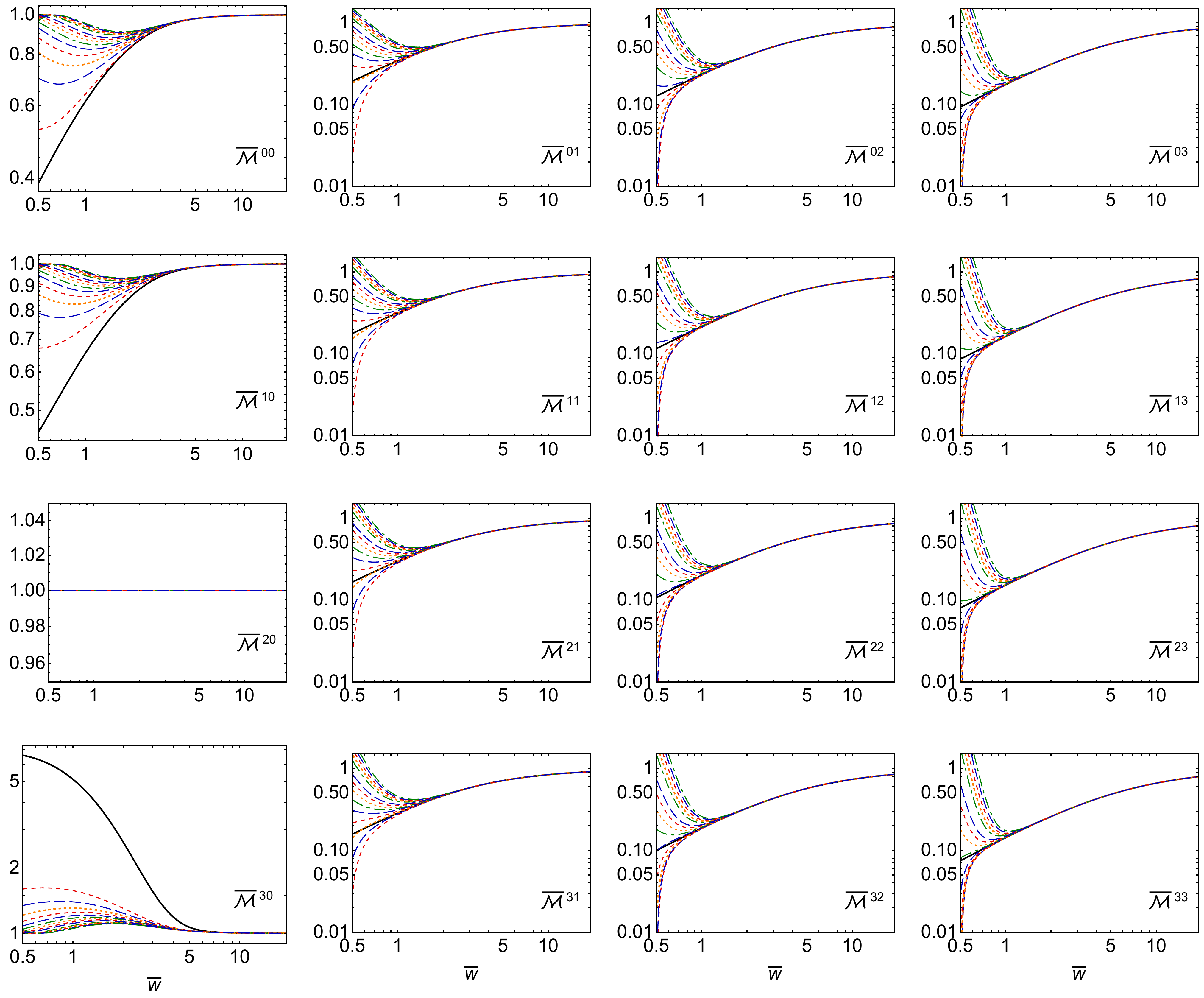}
}
\vspace{2mm}
\caption{Scaled moments $\barM^{nm}$ obtained from the exact attractor solution (solid black line) compared to a set of exact solutions (various colored dotted and dashed lines) initialized at $\tau = 0.1$ fm/c with varying initial pressure anisotropy.  The horizontal axis is $\overline{w} \equiv \tau/\tau_{\rm eq} = \tau T/5 \bar\eta$.  Panels show a grid in $n$ and $m$. }
\label{fig:attractorGridSols}
\end{figure*}

\section{Numerical results}
\label{sect:results}

For a given value of $\bar\eta$ and set of initial conditions specified by $\alpha_0$ and $T_0$, I solve the integral equation \eqref{t4eq} numerically.  For this purpose I wrote a CUDA-based GPU code which allows one to efficiently solve the integral equation \eqref{t4eq} efficiently on very large lattices \cite{MikeCodeDB}.  For all results reported herein, I iterated the integral equation for the temperature \eqref{t4eq} until the result converged to sixteen digits at all values of $\tau$.

In Fig.~\ref{fig:attractorGridSols}, I present the evolution of the scaled moments
\be
\barM^{nm}(\tau) \equiv \frac{{\cal M}^{nm}(\tau)}{{\cal M}^{nm}_{\rm eq}(\tau)} \,,
\ee
associated with the attractor (black solid line) and a set of representative solutions for the   (dashed/dotted colored lines) with differing levels of initial momentum-space anisotropy ($0.1 \leq \alpha_0 \leq 1.5$) and fixed initial temperature \mbox{$T_0 = 1$ GeV} at $\tau_0 = 0.1$ fm/c.  Both the attractor and specific solutions are plotted as a function the scaled time $\overline{w} \equiv \tau/\tau_{\rm eq} = \tau T/5 \bar\eta$.  For the solutions with different initial conditions (dashed/dotted colored lines) I used 2048 points spaced logarithmically between $\tau = 0.1$ and 100 fm/c.  For the attractor solution, I used 4096 points spaced logarithmically between $\tau = 0.001$ and 1000 fm/c and tuned the initial anisotropy to $\alpha_0 \simeq 0.0025$ following a method similar to the one outlined in the appendix of Ref.~\cite{Romatschke:2017vte}.  

As can be seen from Fig.~\ref{fig:attractorGridSols}, all solutions approach the attractor solution in a finite scaled time. The slowest approach is for moments with $m=0$ which appear in the leftmost column of Fig.~\ref{fig:attractorGridSols}.  Considering, for example, $\barM^{30}$, the generic solutions visibly merge with the attractor only after $\overline{w} \gtrsim 6$.  For $m \neq 0$, however, one sees that all moments computed from individual solutions visibly merge with the attractor for $\overline{w} \gtrsim 2$.   

\begin{figure*}[t!]
\centerline{
\includegraphics[width=1\linewidth]{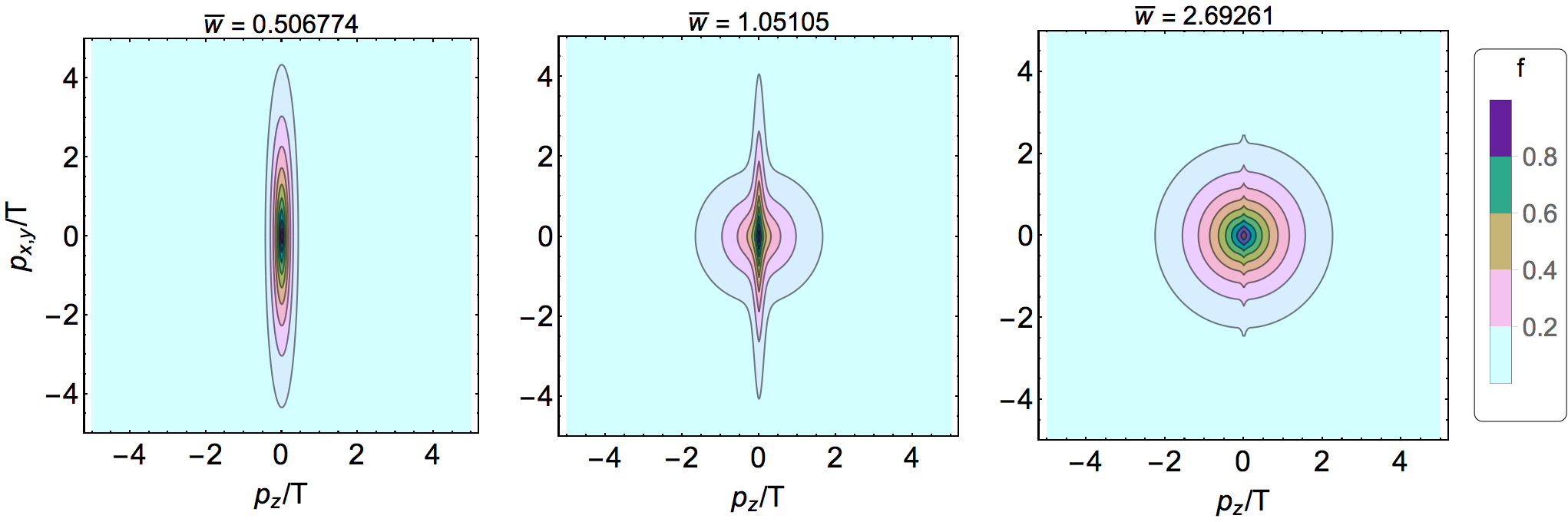}
}
\caption{Visualization of the one-particle distribution function obtained using a typical (non-attractor) anisotropic initial condition.}
\label{fig:f_vis1}
\end{figure*}

In Fig.~\ref{fig:f_vis1} I show visualizations of the one-particle distribution function at three different scaled times. I used a typical anisotropic initial condition from the set shown in Fig.~\ref{fig:attractorGridSols}.  From this figure we see that the exact solution for the one-particle distribution function contains two visually identifiable components.  The first is a highly anisotropic piece which becomes increasingly more compressed into the region with $p_z \sim 0$ as a function of scaled time.  This piece comes from the first term in the exact solution Eq.~\eqref{eq:exactsolf} and corresponds to the free streaming contribution.  As a function of time this contribution becomes more squeezed in the longitudinal direction, however, eventually the amplitude of this very narrow ridge decreases exponentially due to the damping function $D$ in the first term in Eq.~\eqref{eq:exactsolf}.  It is this free-streaming part of the distribution function solution which gives rise to the slower convergence of moments with $m=0$ to their attractors seen in Fig.~\ref{fig:attractorGridSols}.

\section{Conclusions}

In this proceedings, I have summarized the results of an analysis of the exact attractor for the 0+1d RTA Boltzmann equation.  I studied the higher-order moments of the one-particle distribution function and the one-particle distribution function itself.  The results demonstrate that the one-particle distribution exhibits attractor-like behavior in that generic solutions converge to an attractor on a characteristic scaled time scale, which I have dubbed the ``pseudo-thermalization time'', $\overline{w}_c$.  In Ref.~\cite{Strickland:2018ayk}, it was demonstrated that, within RTA, moments with $m>0$ have pseudo-thermalization times that are parametrically shorter than the corresponding thermalization time for $1 \leq m \leq 8$ and $0 \leq n \leq 8$.  This provides explicit proof that there is non-equilibrium attractor that is distinct from the usual late-time Navier-Stokes evolution of the system.

Since Ref.~\cite{Strickland:2018ayk} was published, I have extended the analysis to the case of number-conserving RTA \cite{Strickland:2019hff}.  Therein it was demonstrated that a non-equilibrium attractor also exists, even though the system generically falls out of chemical equilibrium at late times.  Based on this study it would interesting to also look at leading-order scalar field theory, in which case one only has 2 $\leftrightarrow$ 2 collisions and exact number conservation.  Previous works have considered this in the context of aHydro~\cite{Almaalol:2018jmz} and it was shown that one could numerically extract the attractor for both number-conserving RTA and scalar collisional kernels, with the two being qualitatively similar.  One could consider the 2 $\leftrightarrow$ 2 scalar kinetic theory using effective kinetic theory numerical simulations in order to compare with the number-conserving RTA results obtained herein.  It would also be interesting to make comparisons with the attractor extracted from the effective kinetic theory framework of Kurkela et al, particularly in the case that (anti-)quarks are included~\cite{Kurkela:2015qoa,Keegan:2015avk,Kurkela:2018oqw,Kurkela:2018xxd}.

\section*{Acknowledgments}

M. Strickland was supported by the U.S. Department of Energy, Office of Science, Office of Nuclear Physics under Award No. DE-SC0013470.

\bibliographystyle{JHEP}
\bibliography{strickland}

\end{document}